# Spatiotemporal Dynamics of Hydrogen Plasma Smelting Reduction of iron ore: A Multi-Species Diagnostic Approach

Ram Krushna Mohanta,*[a] Kai Lopez,[a] Sai Vishnu Korsipati, [b] Hariswaran Sitaraman, [a] Laxminarayan Raja,[b] Noemi Leick,[c] Seetharaman Sridhar[d]

[a] School for Engineering of Matter, Transport and Energy, Arizona State University, Tempe, AZ, USA, 85281

[b] Department of Aerospace Engineering and Engineering Mechanics, University of Texas at Austin, Austin, TX, USA 78712

[c] National Laboratory of the Rockies, Material and Computational Chemical Science, Golden, CO, USA, 80401

[d] Department of Materials Science and Engineering, Missouri University of Science and Technology, MO, USA, 65409

*Corresponding author: rmohanta@asu.edu, rk21691@gmail.com

Supporting information for this article is given via a link at the end of the document.

## Abstract:

Plasma-based mineral-processing routes, such as hydrogen plasma smelting reduction (HPSR), which converts iron-ore fines directly to liquid metal in a single scalable step are commonly modeled by treating the arc as a spatially uniform heat source. Yet the reduction chemistry is governed by the strongly non-uniform conditions at the plasma–melt interface, which spatially averaged diagnostics cannot resolve. Here we spatially and temporally resolve the arc of a transferred-arc HPSR reactor using multi-species optical emission spectroscopy (OES), in which neutral and ionic argon (Ar I, Ar II), hydrogen Balmer, and neutral iron (Fe I) emissions serve as intrinsic spatial filters set by their differing ionization thresholds. Combined with infrared thermography of the melt surface and an LTE thermal-plasma model validated against the benchmark free-burning argon arc, the measurements reveal a strongly stratified, non-isothermal discharge: an argon-defined core (>10,000 K), a partially recombined Balmer envelope (7,000–10,000 K), and an Fe I-traced interfacial boundary layer (3,000–4,000 K) directly above a melt surface at ~1,900–2,300 K. Across this steep thermal drop, positive hydrogen ions recombine before reaching the surface, so the reductant flux delivered to the oxide is overwhelmingly neutral; atomic hydrogen (H) and vibrationally excited molecular hydrogen $H_2(v)$, rather than the energetic ions often assumed. The measured electron density and excitation temperature bound the interfacial ionization. These findings redefine the boundary conditions for kinetic modeling of plasma-based ore reduction and establish a spatially resolved multi-species diagnostic framework transferable across plasma mineral-processing systems.

**Keywords:** Hydrogen Plasma Smelting Reduction, Thermal plasma, Spectroscopy, Ironmaking

## 1. Introduction

The steel industry faces a dual challenge: meeting unprecedented demand for metallic iron [1–3] while processing increasingly low-grade ores under a shifting energy infrastructure [1]. Among the electrified alternatives to carbothermic reduction, hydrogen plasma smelting reduction (HPSR) is a particularly direct route [4–6]: a thermal arc ionizes the hydrogen-bearing gas and simultaneously reduces and melts iron-ore fines in a single reactor, yielding liquid metal separated from gangue while eliminating the agglomeration steps required by shaft-furnace routes [7,8]. This single-step, fines-to-liquid capability makes the process attractive for scalable primary iron production.

The plasma also transforms the reductant itself. Ionization generates atomic hydrogen (H) and ionic species ($H^+$, $H_2^+$, $H_3^+$) whose free energies of reduction are substantially lower than those of molecular hydrogen[4,6,9,10], while the high arc temperature affords rapid reduction kinetics.

Research into the hydrogen plasma smelting process dates back to the 1980s [11,12], and its mechanisms, kinetics, and potential for sustainable green steel production have been comprehensively reviewed [4,6,13–17]. Recent work has concentrated on optimizing macroscopic operating variables [14,18–24] with optical emission spectroscopy (OES) established as the leading in-situ diagnostic for the process [25–30]. However, existing OES studies predominantly treat the plasma arc as a homogenous macroscopic heat source with uniform chemistry: : broad line-of-sight integration averages properties over the entire arc volume, obscuring the steep thermal gradients and species stratification between the arc core and the melt surface- the localized conditions that actually dictate the reduction chemistry [31].

This research gap is critical because both the thermodynamics and kinetics of reduction, and with them the efficiency of hydrogen utilization depends on which hydrogen species reach the reaction interface and at what concentration. Our previous computational modeling [31] , consistent with established models of thermal arcs [32–34] predicts that the steep gradients and strong electric fields near the anode drive pronounced species separation. Because HPSR operates as a forward-polarity transferred arc, the molten ore itself is the anode, and the resulting near-surface sheath field is expected to repel positive ions ($H^+$), leaving neutral atomic hydrogen (H) and vibrationally excited molecular hydrogen, $H_2(v)$, to dominate the reduction zone. Without spatially resolved measurements, this mechanism has remained a hypothesis, forcing kinetic models to rely on assumed species fluxes that oversimplify interfacial thermodynamics.

Here, we resolve the plasm experimentally. Spatially resolved OES measurements were performed at discrete heights along the arc column, across a range of arc currents, using three distinct emitting species (Ar, H, and Fe) to probe the electron density and temperature of each plasma regime. Combined with infrared thermography of the melt surface and corroborated by computational modeling, these measurements empirically establish the boundary conditions of the reduction zone. The result converts near-anode plasma chemistry from theoretical assumption into verified empirical input, the physical framework required for high-fidelity kinetic modeling and scale-up of hydrogen-based ironmaking.

## 2. Methods and Materials

### 2.1. Experimental set up

The experimental setup (**Figure 1**) and reduction procedure follows the methodology described in detail in our previous work [35]. Reduction experiments were performed at arc currents of 100, 150, and 200 A under a constant 20 L/min flow of an Ar–$H_2$ (95:5 vol.%) mixture, with an electrode gap ($d_{arc}$) of 22 mm. Each experiment used ~10 g of DRI-grade hematite pellets (Voestalpine, USA [36]). The chemical composition of the ore is given in Supplementary information.

Axially resolved emission spectra were acquired with a four-channel Avantes AvaSpec-NXS2048CL spectrometer covering 190–1039 nm (1200 lines/mm gratings, 10 μm slits, ~0.188 nm resolution, 2048-pixel CMOS detectors), with integration times optimized for signal-to-noise ratio. A bi-convex lens (Ø 25.4 mm, f = 100 mm) formed a real, 1.25× magnified image of the arc on the fiber-optic plane, and spatial mapping was performed by mechanically translating the fiber along the arc (z) axis; each spectrum is therefore radially-averaged along the line of sight of the collection optics. Owing to the geometry of the reactor viewport, the accessible range was 16 mm of the 22 mm arc length: z = 0 mm denotes the position immediately above the sample surface (hereafter, the anode), and z = 16 mm the uppermost measured position, 6 mm below the cathode tip. A neutral-density filter in the optical path prevented detector saturation by the intense core emission near the cathode and, because the viewport and lenses were optimized for <0.5% reflectance over 350–700 nm, spectral analysis was restricted to this high-transmission window to ensure radiometric accuracy.

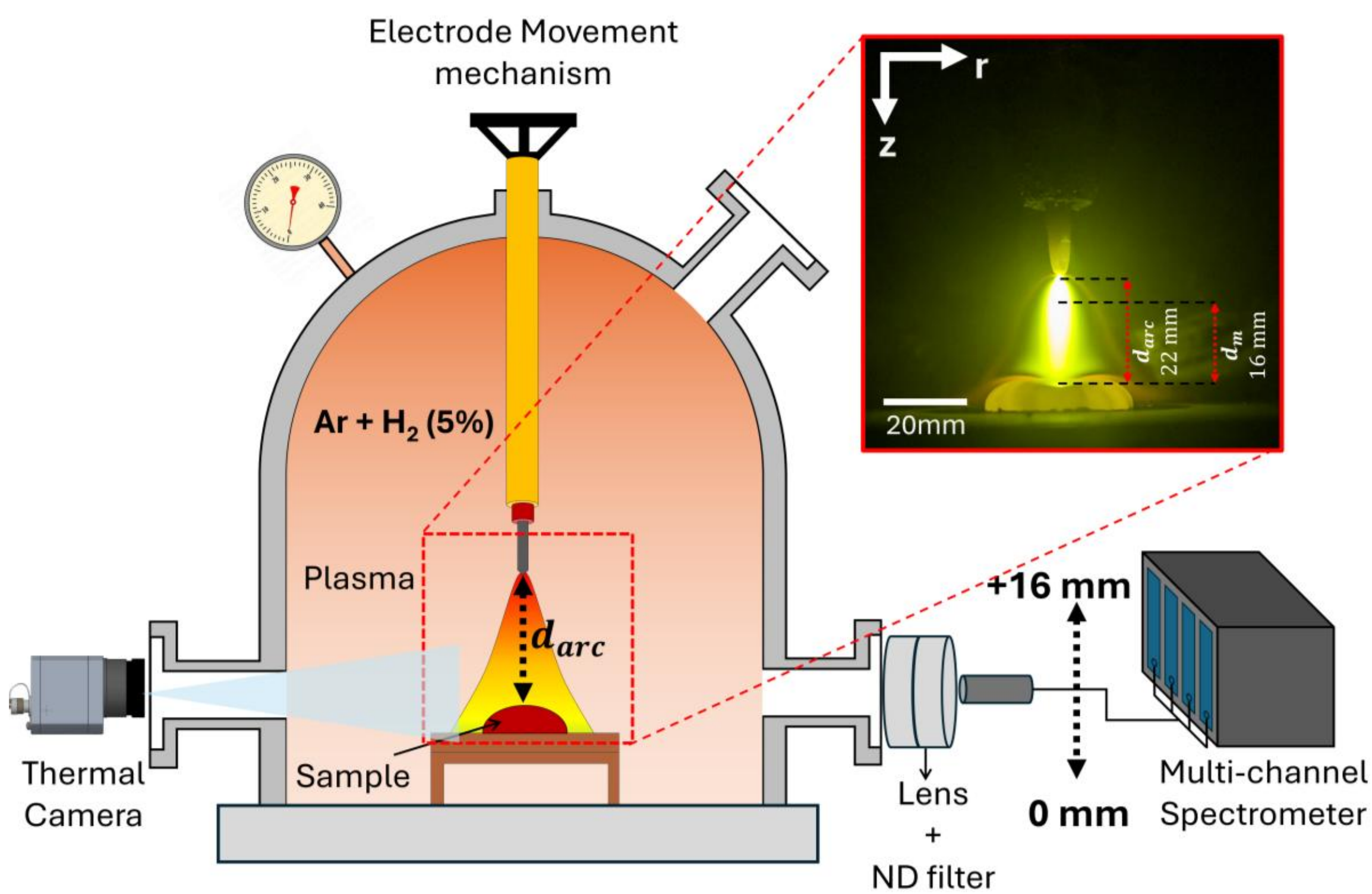

**Figure 1:** Schematic of the experimental set up for HPSR, including the optical emission spectrometer and Infrared camera. Due to geometric constraint of the horizontal view port, the optical fiber can scan 16mm out of 22mm arc length. $d_{arc}$: total arc length, $d_m$: measured arc length.

The sample surface temperature was measured with a short-wavelength thermal imaging camera (Optris PI 05M [37]) operating at 0.50–0.54 μm, a band that minimizes emissivity-related errors on hot, reflective surfaces and permits measurement of molten metals up to 2450 °C. Because the sample transforms during reduction from a dark oxide melt to a reflective metallic pool, the spectral emissivity ε was adjusted accordingly: ε = 0.7 for molten iron oxide and ε = 0.35 [38–40] for fully metallized molten iron.

## 2.2. Computational Method

The computational modeling of thermal plasmas is commonly developed within a fluid dynamics framework, in which the plasma is represented as a viscous, compressible conducting fluid. Under this approach, the governing Navier–Stokes equations are solved in conjunction with Maxwell's equations to describe the coupled fluid flow and electromagnetic behavior of the free-burning arc. The formulation adopted in this study is based on the following assumptions, which are widely accepted in magnetohydrodynamic (MHD) modeling of thermal plasma arcs [34,41–43] :

I. The plasma is modeled as quasi-neutral single fluid medium under local thermodynamic equilibrium (LTE) assumption with plasma species concentrations and transport properties determined by local temperature and pressure.

II. The flow is assumed to remain in the low-Mach number regime with negligible variations in thermodynamic pressure while a perturbation pressure enforces mass conservation.

III. A low magnetic Reynolds number is assumed, implying that the magnetic induction effects are negligible and electric current density is governed predominantly by the applied electric field.

IV. Arc discharge is primarily governed by electrostatic effects as opposed to transient electromagnetic effects, thus allowing time derivatives of electric and magnetic fields to be neglected.

V. The molten oxide surface is assumed to remain flat and stationary throughout the simulation. No-slip boundary conditions are imposed for velocity and temperature at the interface, while free-surface deformation, electrocapillary effects, and other multiphase interactions with the plasma are not considered.

Based on these assumptions, the governing equations consist of the conservation laws for mass, momentum, and energy, coupled with Maxwell's equations describing the electric and magnetic fields, as presented in the Supporting Information. The momentum equation incorporates the Lorentz force as a source term, while the energy equation accounts for Joule heating and radiative losses. The fidelity of the computational approach was validated against the canonical 200 A free-burning argon arc of Hsu et al.[34] whose measured isotherms are in good agreement with this model's prediction (Fig. S4).

# 3. Results and discussion

## 3.1. Spatial and Temporal Distribution of Electron Density ($n_e$)

**Figure 2** presents the spatial distribution and temporal evolution of $n_e$ at 100, 150, and 200 A. Higher arc currents sustain higher overall densities: peak values rise from ~5.8 × $10^{22}$ $m^{-3}$ at 100 A to ~1.2 × $10^{23}$ $m^{-3}$ at 200 A and reflecting the greater ionization required to carry the discharge current. Along the axis, $n_e$ increases monotonically toward the cathode: at 200 A it rises nearly four-fold over the 16 mm measured span, from ~3.1 × $10^{22}$ $m^{-3}$ at the anode interface to ~1.2 × $10^{23}$ $m^{-3}$ at z = 16 mm.

This gradient follows directly from current continuity in the expanding arc column. As the plasma expands from the constricted cathode spot toward the broad molten anode, the conducting cross-section (A) increases and the current density ($J = I/A$) must fall; since $J \propto n_e e v_d$, the plasma becomes correspondingly more diffuse near the anode. Conversely, the severe constriction at the cathode tip demands a high local current density and a dense flux of positive ions to sustain thermionic emission.

The low density at z = 0 mm thus reflects both macroscopic arc expansion into a diffuse, bell-shaped attachment and severe thermal quenching by the comparatively cool melt: the oxide melt interacts with a diffuse plasma fringe rather than the high-energy core. This diffuse attachment is advantageous for HPSR, spreading the thermal flux and preventing localized boiling of the bath while maintaining a reactive environment. Furthermore, quasi-neutrality requires the electron depletion in this boundary layer to be matched by an equivalent drop in the total positive-ion density; because the feed gas is 95% Ar, $n_e$ tracks the quenching of bulk $Ar^+$ rather than the $H^+$ population. Hydrogen speciation is instead set by the interfacial temperature: the steep thermal drop drives the Saha balance toward recombination, so that the hydrogen reaching the oxide surface is overwhelmingly present as reactive neutral species, atomic H and $H_2$(v), rather than $H^+$, identifying these neutrals as the primary reducing agents at the plasma–oxide interface.

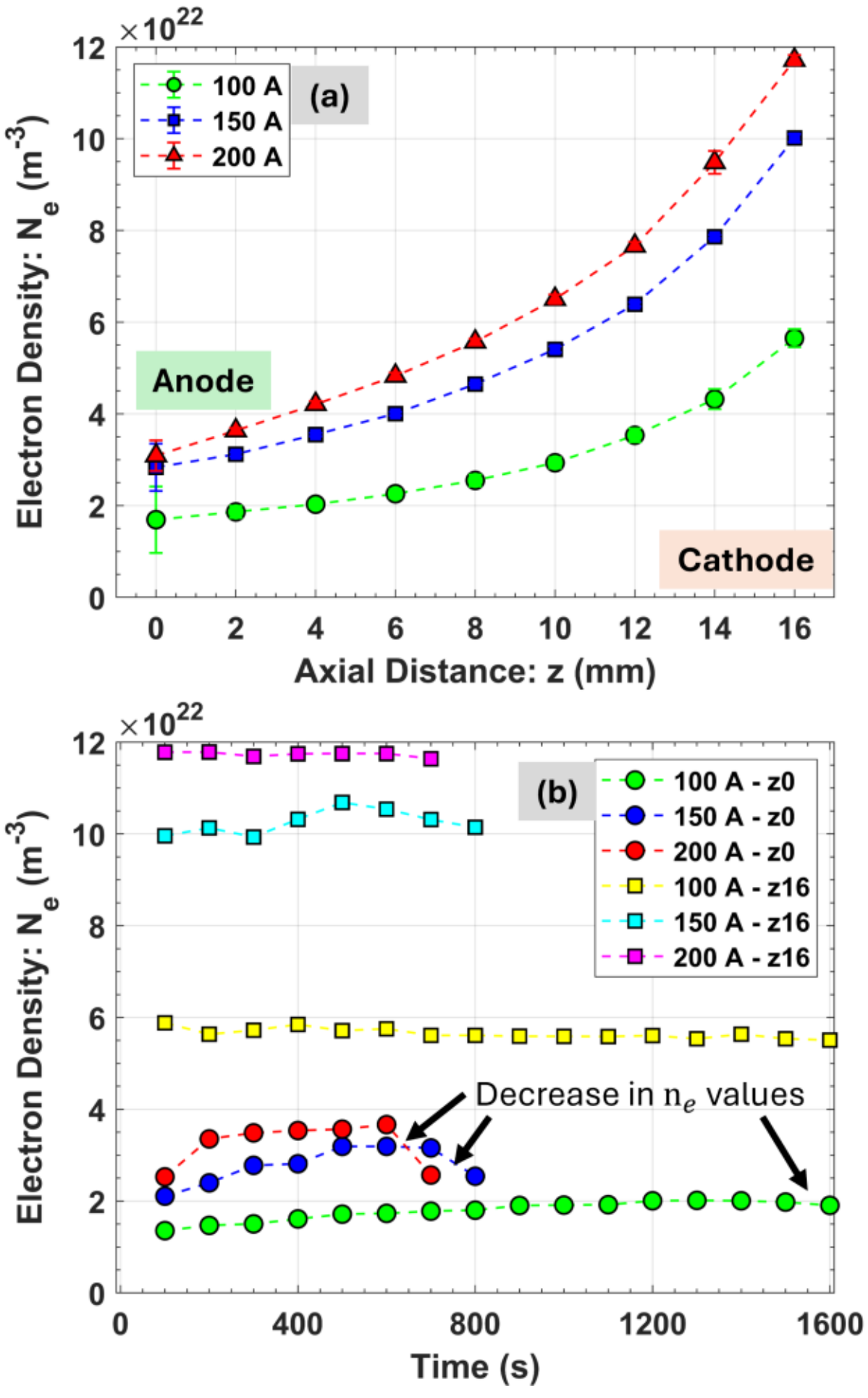

**Figure 2:** Experimental spatial and temporal evolution of $n_e$ (a) Time-averaged axial profiles of $n_e$ for 100, 150 and 200 A arc currents. (b) Temporal evolution of $n_e$ measured at z=0 and z=16mm; arrows mark the onset of the near-anode density drop coinciding with the final stage of reduction, when metallic Fe is exposed to the arc. All values are radially-averaged along the line of sight. z = 0 mm denotes the position ~1 mm above the melt surface; z = 16 mm lies 6 mm below the cathode tip; total arc length is 22 mm.

The temporal variation of $n_e$ at different axial positions is shown in **Figure 2(b)**. The electron density $n_e$ near the cathode (z = 16 mm) is stable throughout reduction, indicating that thermionic emission decouples the cathode boundary from the evolving melt chemistry below. Near the melt surface, by contrast, $n_e$ is highly sensitive to reduction progress: after an initial gradual rise, pronounced drops appear at ~600 s (200 A) and ~700 s (150 A), coinciding with metallization degrees exceeding ~95% [35], whereas the 100 A profile remains relatively stable over the full measurement (>1400 s).

Because HPSR mechanism is an interfacial reaction, reduction proceeds unevenly from the top of the slag: the surface Fe/FeO ratio runs well ahead of the bulk, while density-driven separation drains metal to the bottom. In the final stages, local depletion of the resistive oxide layer exposes metallic Fe directly to the arc, forming a highly conductive anode spot and triggering intense Fe evaporation. The resulting vapor influx abruptly alters the local plasma state: although iron's low ionization potential (7.9 eV vs. 15.76 eV for Ar) would nominally enhance ionization, Fe is a strong radiator, and radiative cooling of the anode boundary layer lowers the local temperature enough to suppress the net ionization rate, thus producing the density drops marked by arrows in **Figure 2(b)**. These drops occur only at 150 and 200 A suggests the underlying Fe evaporation requires a threshold energy flux reached only at higher currents.

## 3.2. Spatial and Temporal Distribution of Plasma Excitation Temperature ($T_{exc}$)

The measured excitation temperature depends strongly on the emitting species used to derive it which is a direct signature of the arc's non-uniform thermochemical structure. $T_{exc}$ obtained from Ar I and Ar II transitions via the Saha–Boltzmann method is the highest, 10,000–13,000 K. Because $Ar^+$ formation requires 15.76 eV, its population is heavily biased toward the high-electron-temperature core, where $n_e$ peaks and LTE is most closely approached; although each measurement is radially-averaged, the signal is dominated by this hot core, making Ar-based diagnostics a reliable proxy for the core temperature.

As illustrated in **Figure 3(a)**, the $T_{exc}$ increases monotonically with axial distance from anode to cathode at all currents: at 200 A, $T_{exc}$ rises from $11{,}400 \pm 350$ K at the anode interface (z=0) to a peak of $13{,}100 \pm 400$ K near the cathode (z=16), representing a thermal gradient of ~1700 K across the gap. Even at the lowest current of 100 A, a gradient of ~1000 K is maintained (10,600 K to 11,600 K). This axial variation reflects the electrode boundary conditions: intense Joule heating at the constricted thermionic cathode spot sustains the maximum temperature, whereas the molten oxide anode acts as a heat sink, dissipating energy through conduction into the melt, the endothermic reduction reaction, and the latent heat of vaporization. This cooling is central to HPSR as the reductant species reaching the ore are markedly colder than the arc core, favoring recombination of hydrogen ions near the surface.

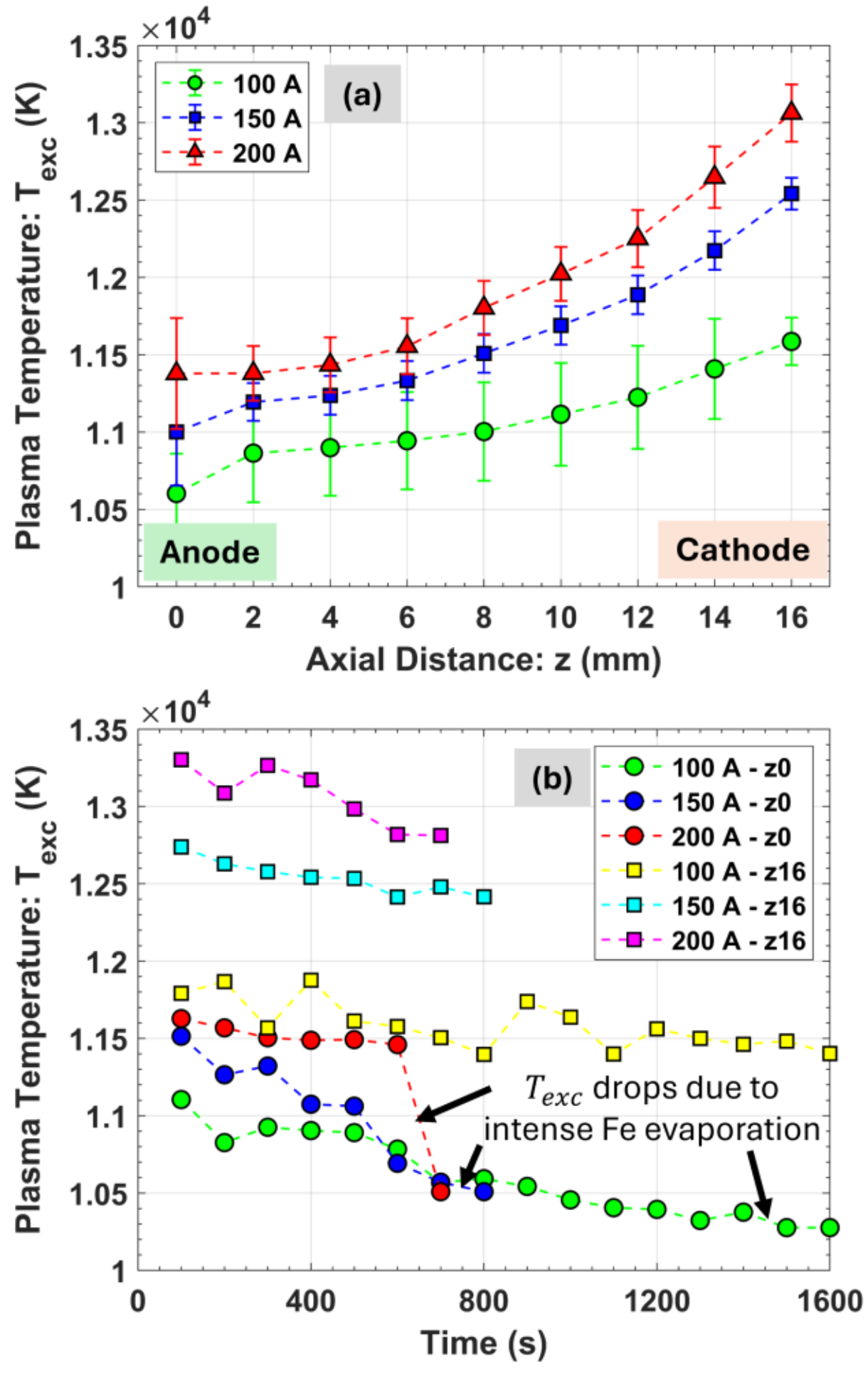

**Figure 3:** Experimental spatial and temporal evolution of $T_{exc}$ during HPSR (a) Time-averaged axial profiles of $T_{exc}$ for 100, 150 and 200 A arc currents. (b) Temporal evolution of $T_{exc}$ measured at z=0 and z=16mm: arrows mark the onset of the temperature drop at z = 0 mm coinciding with the final stage of reduction, when metallic Fe is exposed to the arc. All values are radially-averaged along the line of sight. z = 0 mm

denotes the position ~1 mm above the melt surface; z = 16 mm lies 6 mm below the cathode tip; total arc length is 22 mm.

The temporal evolution of the $T_{exc}$ at z=0, shown in **Figure 3(b)**, shows the anode-side temperature declines gradually as reduction proceeds and then collapses at the endpoint. At 200 A, $T_{exc}$ falls from ~11,800 K through a sharp ~1,000 K drop at ~600 s, stabilizing near 10,500 K. The drop coincides with the exposure of metallic Fe (metallization > 95%) and attributed to the Fe-vapor mechanism explained in Section 3.1, where increased Fe evaporation in plasma enhances radiative cooling and increases plasma thermal conductivity.

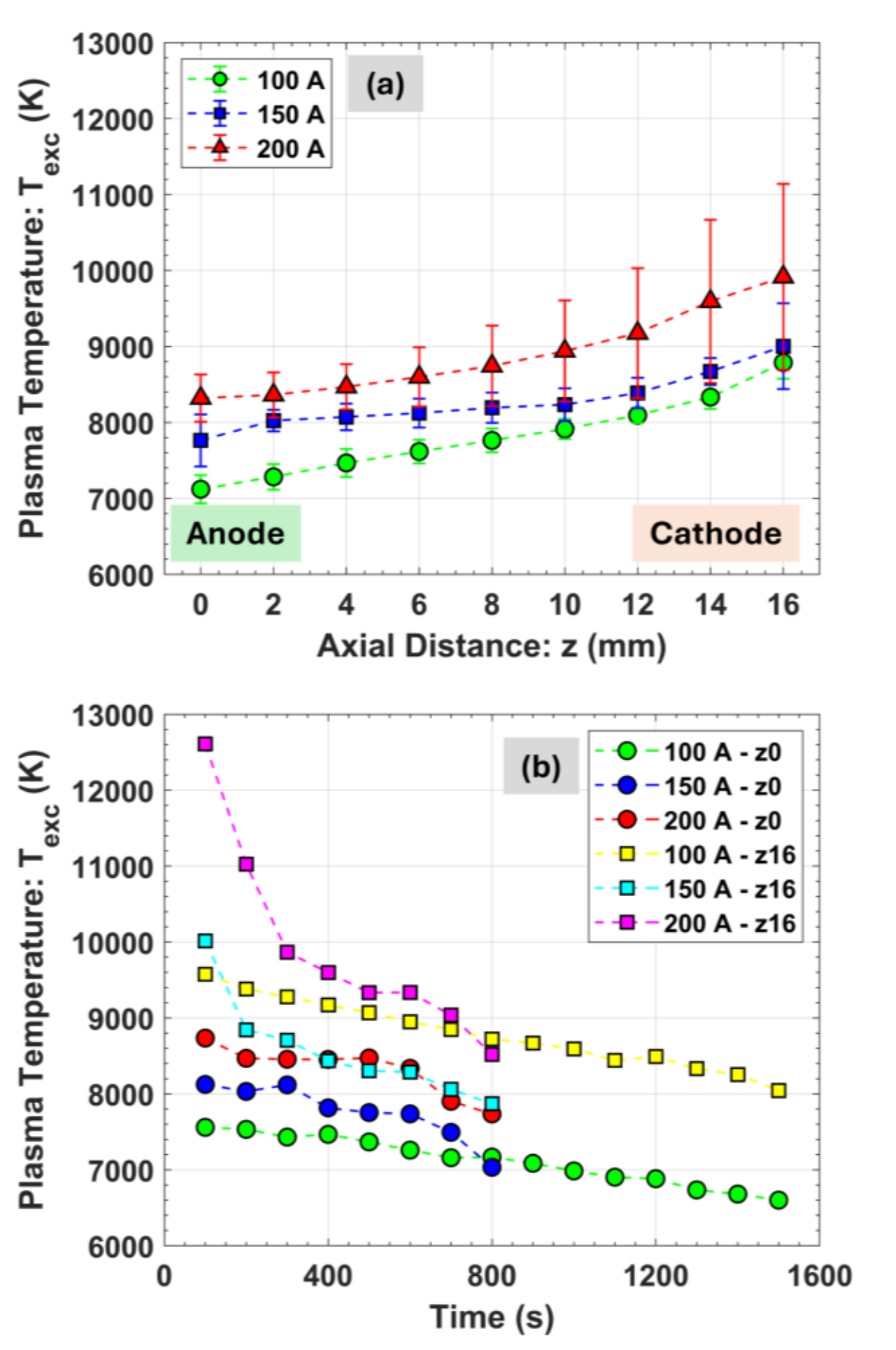

**Figure 4:** Experimental spatial and temporal evolution of $T_{exc}$ calculated using Balmer series Hydrogen lines($H_\alpha$, $H_\beta$). (a) Time averaged axial profiles of $T_{exc}$ for 100, 150 and 200 A arc currents (b) Temporal variation of $T_{exc}$ at anode surface (z=0mm) and near cathode (z=16mm).

The Balmer-derived $T_{exc}$ is consistently lower, spanning 7,000–10,000 K. Hydrogen emission in this system is fed by electron-impact excitation, $H_2$ dissociation, and radiative-recombination cascades ($H^+ + e^- \rightarrow H^*$); because recombination is most active in the cooler intermediate and fringe regions, where atomic hydrogen is abundant but the temperature has fallen below core levels, Balmer series emissions ($H_\alpha$ and $H_\beta$) inherently probe a partially recombined envelope rather than the core.

In **Figure 4** (a), the hydrogen-derived profiles of axial variations of $T_{exc}$ mirror the Ar trend; rising toward the cathode to a maximum of ~10,000 K at 200 A under the same electrode boundary conditions but sit 2,000–3,000 K below the Ar-derived values, confirming that Balmer series emission selectively samples the cooler envelope of the expanding plume. In **Figure 4** (b), temporal variations of $T_{exc}$, the anode-side temperature declines steadily over each run: ~8,800 to ~7,800 K at 200 A (by ~600 s), ~8,100 to ~7,100 K at 150 A, and ~7,500 to ~6,600 K at 100 A. The near-cathode position shows a sharper transient, plummeting from a brief initial peak of ~12,600 K to ~9,000 K at 200 A as the arc expands. The steady anode-side cooling directly tracks the accumulation of evaporated Fe in the boundary layer as metallization progresses.

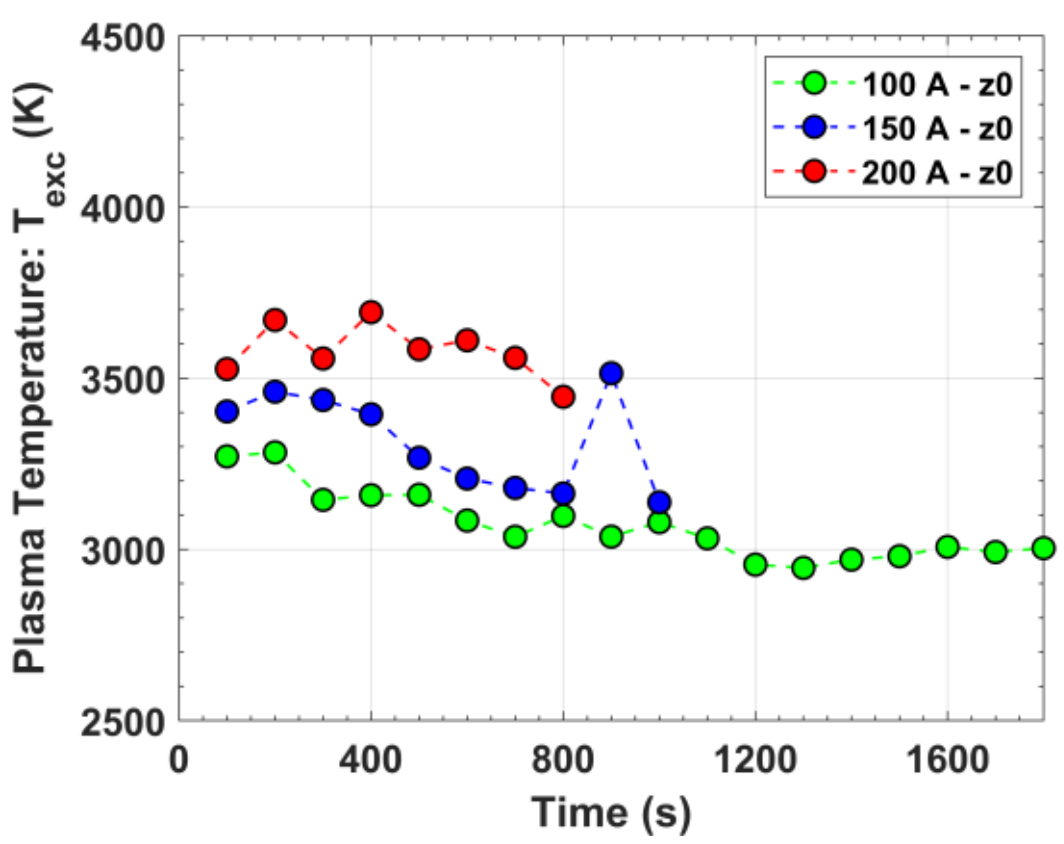

**Figure 5:** Experimental temporal evolution of $T_{exc}$ derived from Fe I lines at the anode surface (z = 0 mm) for 100, 150, and 200 A. The measured 3,000–4,000 K values reflect the thermal boundary layer at the plasma–oxide interaction zone.

Fe I emission at the anode surface yields far lower values, 3,000–4,000 K (**Figure 5**), with a gradual decline mirroring the reduction: ~3,500–3,700 K with a small terminal drop at 200 A,

~3,400 to ~3,100 K at 150 A, and ~3,300 to a stable ~3,000 K at 100 A. These values reflect the plasma–liquid boundary layer, typically 0.1–0.5 mm thick [31,44,45]. Whereas the Ar and Balmer transitions have upper-level excitation energies above 12 eV and therefore light up only in the hot gas envelope, Fe I emission is confined to the interface by two effects: the vapor originates at the evaporating melt surface, and iron's low ionization threshold (~7.9 eV) ensures that any Fe diffusing into the >8,000 K core is rapidly ionized to $Fe^{+}$. Strong local cooling - conduction into the melt, latent heat of evaporation, and radiative losses enhanced by the iron vapor itself - suppresses the boundary-layer temperature, so Fe I diagnostics measure the plasma directly at the reaction surface. By the same token, Fe I lines cannot probe the upper arc: the neutral population is concentrated near the anode, and upward-diffusing vapor is almost entirely ionized in the >10,000 K core. The resulting collapse in Fe I intensity introduces large statistical errors into the Boltzmann fit, rendering Fe I-derived $T_{\mathrm{exc}}$ unreliable in the near-cathode regime.

### 3.3. Implications for hydrogen plasma reduction chemistry

The stratified temperatures directly identify the active reducing species at the reaction interface. Standard LTE treatments assign a single bulk temperature to the entire plasma–melt system; the multi-species diagnostics show that the interfacial thermodynamics are fundamentally decoupled from the arc core. Because each OES measurement is radially-averaged, the emitting species act as natural spatial filters governed by their ionization thresholds, and the thermal structure of the HPSR plasma resolves into four regimes (**Figure 6**):

a) **Region 1:** The Hot Core (>10,000 K): Governed by argon's high ionization threshold (15.76 eV), this quasi-LTE zone is the primary electrical and thermal energy source, dominated by ionic emission.
b) **Region 2:** The Intermediate Zone (7,000-10,000 K): Probed by the hydrogen Balmer series (H ionization threshold 13.6 eV); an active transport and partial-recombination zone spanning the steep thermal gradients.

c) **Region 3:** The Fringe Region: A diffuse transitional buffer cooling radially toward the ambient gas.

d) **Region 4:** The Plasma-Oxide Interface (3,000-4,000 K):Traced exclusively by neutral iron (Fe I), owing to its comparatively low ionization threshold (~7.9 eV) relative to H and Ar. This ionization-energy logic is not specific to Fe: Mn (7.43 eV) and Mg (7.65 eV), common constituents of lower-grade ores and slags, have nearly identical thresholds, and their neutral vapors would likewise be confined to the cool interfacial layer. Mn I emission (~403 nm) could therefore serve equally for multi-line Boltzmann thermometry of the reaction surface, while the Mg I triplet (~517 nm) offers a complementary spatial and abundance tracer extending the present diagnostic to a broad range of industrial feedstocks.

The ultimate chemical implication lies in Region 4. Because metallurgical reduction occurs exclusively at the liquid surface, the state of the hydrogen reactants is dictated by this 3,000–4,000 K boundary layer rather than by the arc core. Severe thermal quenching forces rapid recombination of descending $Ar^+$ and $H^+$, so the oxide interface is not bombarded by energetic ions but exposed to an overwhelming gas-phase flux of reactive neutrals (atomic H and $H_2$(v)). Dissociation of $H_2$ remains substantial within this 3,000–4,000 K layer, and because the gas number density rises as the temperature falls at constant pressure, these reactive neutrals are further concentrated toward the surface. The macroscopic reduction environment is therefore governed by the neutral pathways $FeO + 2H / H_2(v) \rightarrow Fe + H_2O$, rather than by the previously assumed high-energy ionic reactions.

While this quenching renders the interfacial $H^+$ density orders of magnitude below that of the reactive neutrals, the reduction kinetics also depend on surface adsorption. Preliminary surface-kinetic investigations suggest $H^+$ may exhibit a substantially higher sticking coefficient than the neutral species, so the disproportionately high surface reactivity of the surviving trace $H^+$ cannot be entirely discounted. The interplay between the abundant, lower-sticking $H/H_2$(v) and the scarce, higher-sticking $H^+$ remains a key target for future kinetic modeling, especially since the boundary layer likely departs from thermal equilibrium, with the electron temperature exceeding the gas temperature, rendering any single-temperature treatment of the interfacial ionization balance approximate. Nevertheless, the measured collapse of electron density and excitation temperature establishes the interface as a severely quenched, weakly ionized boundary layer: accurate kinetic models can no longer rely on bulk-plasma $H^+$ concentrations.

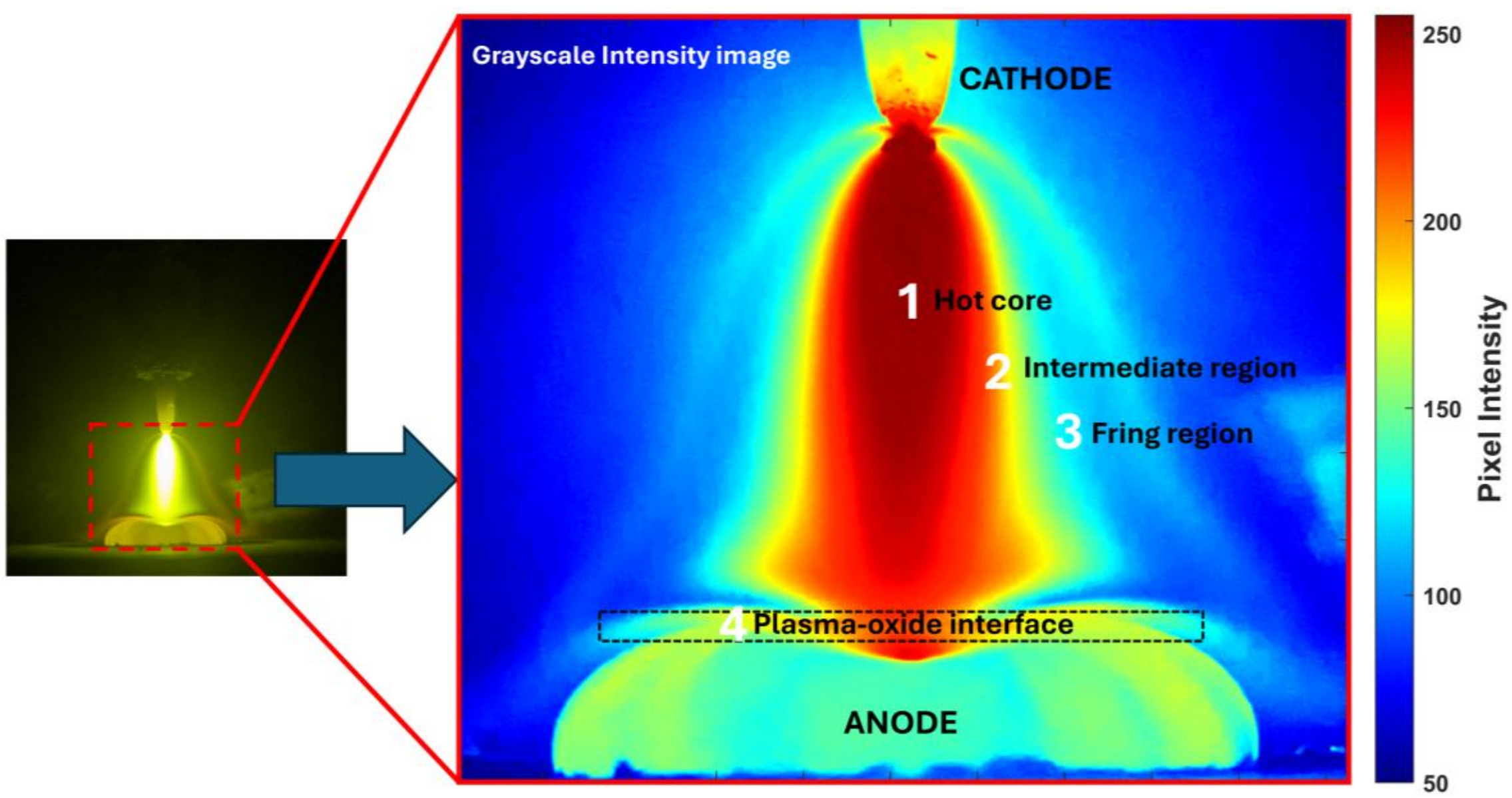

**Figure 6**: Grayscale intensity map of the plasma illustrating its spatial stratification into four regions: (1) hot arc core, (2) intermediate zone, (3) fringe, and (4) plasma–oxide interface. The regions correspond to the temperatures inferred from different emitting species, demonstrating the multi-scale thermal structure of the HPSR plasma.

Although the present experiments were limited to 5% $H_2$ by the operational safety envelope, the method applies at any $H_2$ concentration: the spatial ordering is fixed by the ionization-energy hierarchy of the emitting species (Fe 7.90 eV < H 13.6 eV < Ar 15.76 eV) and by the thermal quench at the melt boundary, not by feed-gas composition. A higher $H_2$ fraction constricts the arc [46,47], steepening the gradients and sharpening rather than blurring the spatial separation.

## 3.4. Melt-surface temperature from infrared thermography

While OES resolves the thermal state of the plasma and its gas-phase boundary layer, completing the heat-transfer picture requires measurement of the condensed-phase anode. Throughout most of the reduction, the sample adopts a core–shell ("egg-shell") morphology, confirmed by cross-sections of interrupted experiments (**Figure 7** d–f), the denser, newly reduced iron coalesces at the bottom and core of the crucible, while the lighter unreduced oxide forms a continuous outer layer directly exposed to the arc. The sample geometry mirrors this progression **Figure 7** a-c: at t = 100 s the oxide melt wets the crucible as a wide, flattened dome; by t = 500 s the oxide shell encapsulates a growing metallic core; and by t = 1000 s the shell is depleted and the fully metallized sample contracts into a compact, near-spherical pool under its altered surface tension and density. Because the camera therefore views the oxide layer for the bulk of the run, ε = 0.7 was applied throughout, switching to ε = 0.35 only beyond ~95% metallization, when the reflective iron pool is exposed.

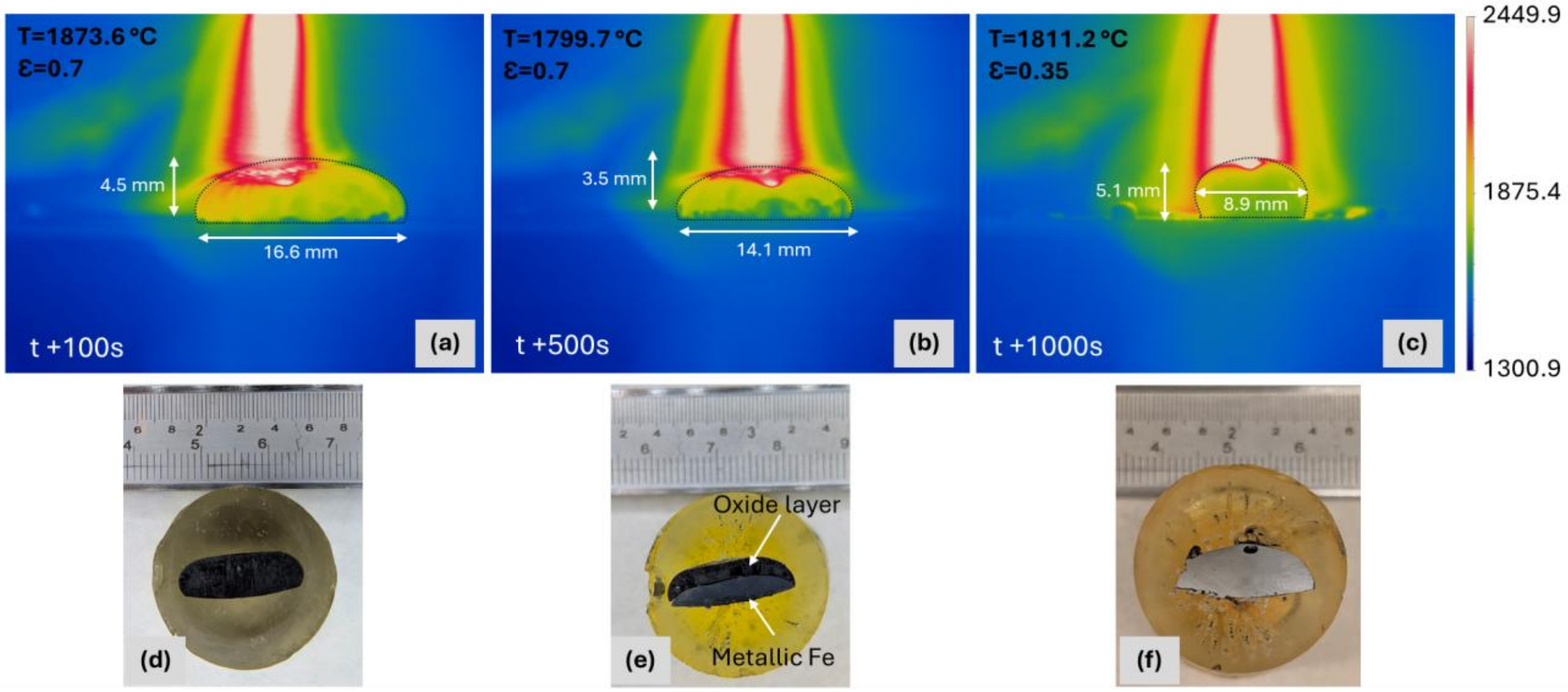

**Figure 7:** Temporal evolution of the sample morphology during HPSR at 150 A. (a–c) Infrared thermal images at t = 100, 500, and 1000 s, showing the geometric transition from a wide oxide melt to a compact metallic sphere. (d–f) Corresponding macroscopic cross-sections of interrupted samples, confirming the core–shell morphology in which the denser metallic core is encapsulated by the lighter unreduced oxide layer.

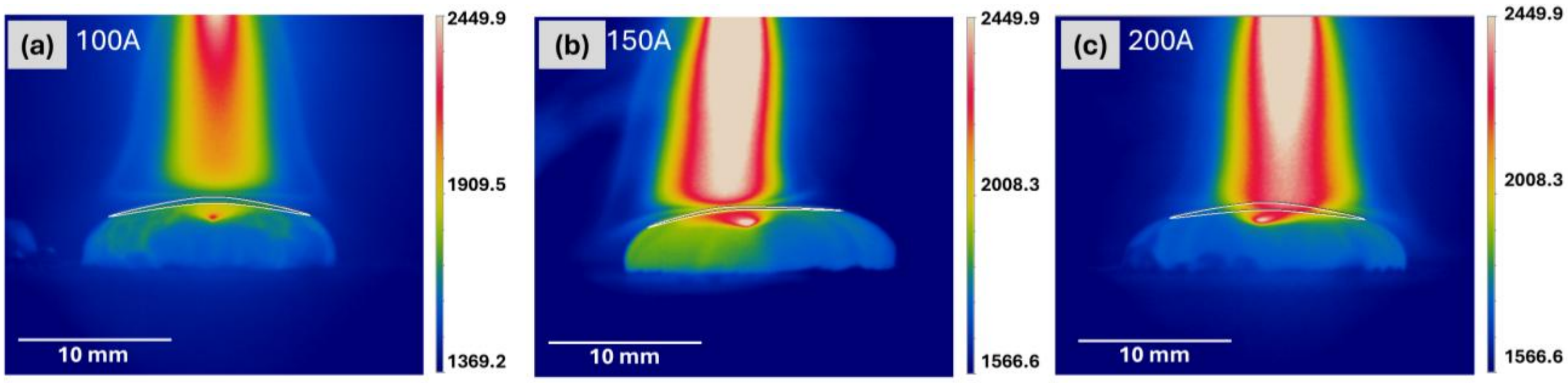

**Figure 8:** Infrared thermal images of the molten oxide surface during steady-state reduction at (a) 100 A, (b) 150 A, and (c) 200 A. The outlined region denotes the primary plasma–anode attachment zone used for temperature averaging; ε = 0.7, corresponding to the molten oxide layer encapsulating the metallic core during the primary reduction stages.

With the radiometric parameters so defined, the slag surface temperature was evaluated during the steady-state reduction phase (**Figure 8**). Within the region of interest at the primary plasma–anode attachment zone, the time-averaged surface temperature rises with arc current - from ~1,940 K (1,670 °C) at 100 A to ~2,210 K (1,940 °C) at 150 A and ~2,310 K (2,040 °C) at 200 A - reflecting the enhanced Joule heating and heat transfer from the expanding plume.

Coupling these surface temperatures with the OES diagnostics completes the thermal map of the plasma–liquid interface: the temperature cascades from >10,000 K in the arc core, through 3,000–4,000 K in the Fe I-probed gas boundary layer, to ~1,900–2,300 K at the liquid surface. This steep cascade quantifies the interfacial thermal boundary layer: the immense heat flux delivered by the arc is consumed at the surface by the latent heat of vaporization, conduction into the water-cooled crucible, and endothermic reduction reactions.

## 3.5. Computational results

The transferred-arc configuration - ore surface as anode, tungsten pin as cathode - was simulated at the experimental current levels using the thermal plasma model developed in our previous work [31]. The fidelity of this computational approach was first established by benchmarking against the canonical 200 A free-burning argon arc of Hsu et al.[34], whose measured isotherms the model reproduces with good agreement (Fig. S4, Supplementary Information), lending confidence to its extension to the Ar–$H_2$ system studied here. Because the stiffness of the coupled governing equations severely constrains the allowable time step, the simulations do not track the full reduction cycle, over which the temperature varies by a few hundred kelvins. A free-burning arc, however, reaches a quasi-steady state within tens of milliseconds - stabilized arc column, developed electrode sheaths, balanced energy transport - while reduction proceeds over hundreds of seconds. Exploiting this timescale separation, the simulations resolve the quasi-steady arc representative of the initial phase of the reduction cycle.

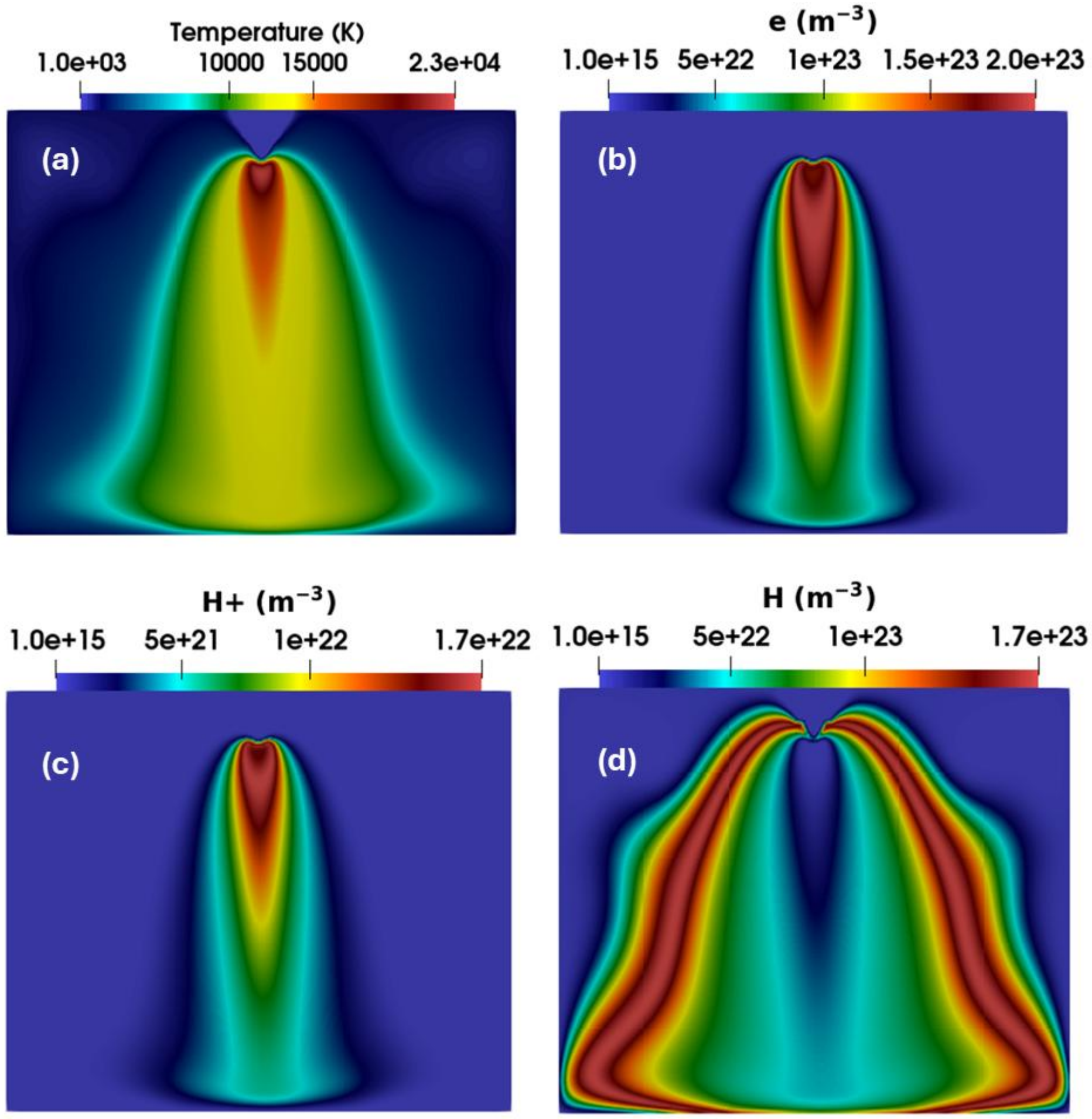

**Figure 9:** Computed mid-plane distributions for a 200 A arc discharge [31]: (a) plasma temperature, (b) electron density, (c) $H^+$ ion density, and (d) neutral H density.

The computed temperature field (**Figure 9**(a), mid-plane slice) peaks at ~2 eV (1 eV = 11,604 K) near the cathode and becomes progressively more diffuse toward the anode, reproducing the characteristic bell-shaped discharge profile and diffuse anode attachment observed in the IR images (**Figure 8**). The corresponding LTE species fields (**Figure 9**b–d) show that hydrogen speciation is governed by the local temperature: electrons and $H^+$ concentrate in the 1–2 eV core, while neutral atomic hydrogen dominates the cooler periphery. Molecular hydrogen dissociates efficiently at ~0.3–0.6 eV [48,49] and the resulting atoms are progressively ionized as the temperature approaches 1–2 eV, confining elevated $H^+$ densities to the hot central column. The model thus independently reproduces the thermochemical stratification observed spectroscopically - a hot ionized core enveloped by a neutral-dominated periphery - corroborating the species-filtering interpretation of the multi-species OES measurements.

## 4. Conclusion

This study used spatially resolved multi-species OES and infrared thermography to dismantle the single-temperature description of the HPSR process, mapping the localized thermodynamic conditions at the plasma–melt interface. Three conclusions follow:

i. **Plasma Stratification and Thermal Decoupling:** The HPSR plasma is strongly stratified, rendering bulk LTE assumptions insufficient. Using the distinct ionization thresholds of Ar, H, and Fe as natural spatial filters, we showed that the >10,000 K arc core is fundamentally decoupled from the chemistry at the melt interface.
ii. **Interfacial Boundary Layer:** Fe I emission and infrared thermography place the reaction surface beneath a steep thermal boundary layer: 3,000–4,000 K in the immediate gas phase, falling to ~1,900–2,300 K at the melt surface. Together with the diffuse, bell-shaped anode attachment, this distributes the thermal flux and prevents localized boiling of the oxide bath.
iii. **Neutral Reduction Pathway:** Thermal quenching within this boundary layer dictates the chemical state of the reductant: recombination of descending ions ensures that the gas-phase flux reaching the liquid oxide is overwhelmingly composed of reactive neutrals, atomic H and $H_2(v)$, establishing a neutral-pathway reduction mechanism in place of the previously assumed ionic reactions.

Ultimately, this thermal decoupling explains the physical stability of the HPSR process and redefines its kinetic framework. Because boundary-layer quenching is intrinsic to plasma–melt interactions, the neutral reduction pathway identified here extends to a wide variety of iron ores, including Mn- and Mg-bearing feedstocks, whose low-ionization-energy neutral vapors are amenable to the same spatially resolved OES analysis and to any higher $H_2$ concentrations, where arc constriction only sharpens the species stratification. Future high-fidelity kinetic models must now integrate surface-adsorption parameters, such as the sticking coefficients of surviving trace ions; the empirically derived boundary conditions reported here provide the physical foundation for such modeling and for the scale-up of plasma-based ore reduction.

## Author Contribution

**RKM:** Writing – original draft, Experimental Investigation-lead. **KL:** Experimental Investigation-support. **SVK:** Computation-support, Writing-review & editing. **HS:** Computation-lead, Writing – review & editing, Project administration, Funding acquisition. **LR:** Writing – review & editing, Supervision, Project administration, Funding acquisition, **NL:** Writing – review & editing, Project administration, Funding acquisition. **SS:** Writing – review & editing, Supervision, Project administration, Funding acquisition.

## Acknowledgements

This work was supported by the U.S. Department of Energy (DOE), Office of Science, Office of Basic Energy Sciences (BES), Materials Sciences and Engineering Division under Award DE-SC0024724 "*Fundamental Studies of Hydrogen Arc Plasmas for High-efficiency and Carbon-free Steelmaking*". This work was authored in part by National Laboratory of the Rockies, operated under Contract No. DE-AC36-08GO28308. The views expressed in the article do not necessarily represent the views of the DOE or the U.S. Government. The U.S. Government retains and the publisher, by accepting the article for publication, acknowledges that the U.S. Government retains a nonexclusive, paid-up, irrevocable, worldwide license to publish or reproduce the published form of this work, or allow others to do so, for U.S. Government purposes.

## Author declarations

The authors have no conflicts to disclose.

# Supplementary Information

## 1. Experimental Methods-Supplementary information

### 1.1. Stark broadening method for electron density calculation

In atmospheric pressure thermal plasmas, the electron density ($n_e$) is a critical parameter governing transport properties and reaction kinetics. The most reliable method for determining $n_e$ in hydrogen-containing plasmas is the analysis of the Stark broadening of Hydrogen Balmer series lines, specifically the $H_\beta$ line (486.1 nm) [1,2].

Stark broadening occurs due to the interaction of the emitting atoms with the electric fields generated by surrounding charged particles (electrons and ions). This perturbation splits and shifts the atomic energy levels, resulting in a broadening of the spectral emission line. Among the Balmer series, the $H_\beta$ line is preferred for diagnostic purposes because it exhibits a strong linear Stark effect, is sensitive to electron density changes, and is less susceptible to self-absorption compared to the $H_\alpha$ line [3]. In the high-density regime typical of atmospheric arcs ($n_e > 10^{22} m^{-3}$) Stark broadening dominates the spectral profile. Other broadening mechanisms, such as Doppler broadening (due to thermal motion) and instrumental broadening, are comparatively negligible (typically < 0.2 nm) against the Stark width (several nanometers) [4]. Consequently, the relationship between the Full Width at Half Maximum (FWHM) of the $H_\beta$ line and the electron density is largely independent of electron temperature, making it a robust standalone diagnostic.

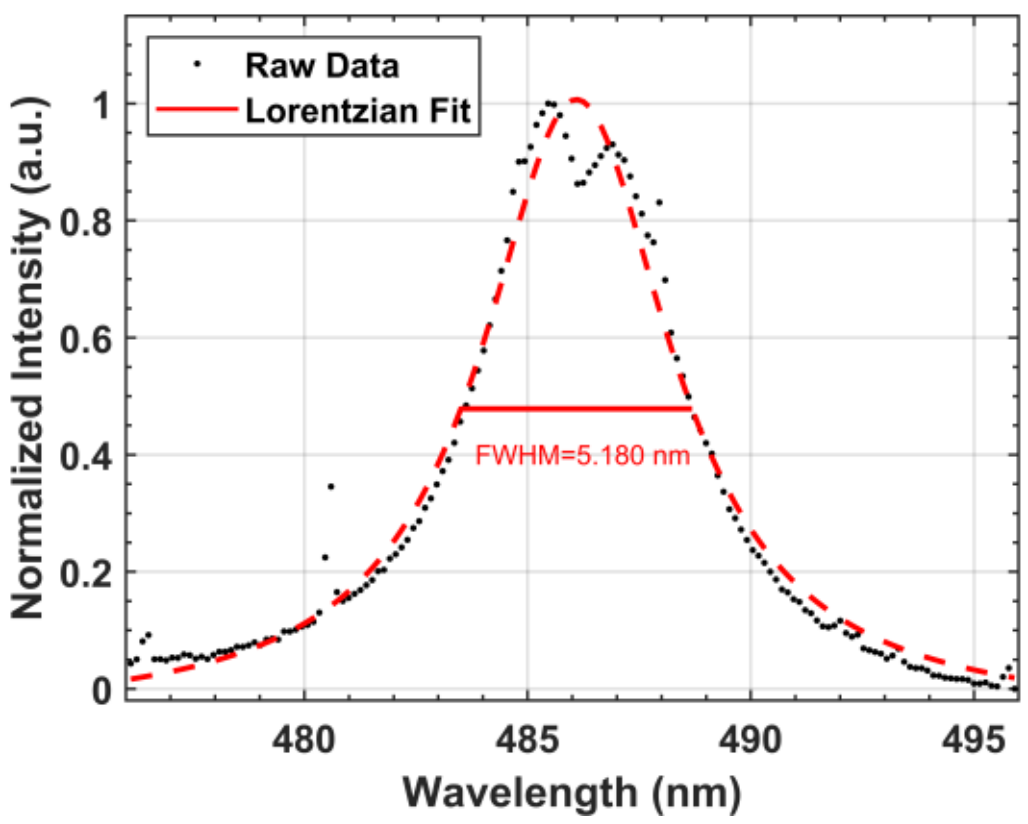

**Figure S1**: An example of the Lorentzian fit to the $H_\beta$ emission spectrum, I=200A. The straight horizontal line indicates the FWHM value

In this study, electron density was calculated using a custom MATLAB algorithm based on the parameterization by Gigosos and Cardenoso [5]. The processing pipeline for each spectrum followed these steps:

1. Line Isolation: The spectral window was restricted to a ±5 nm window to fully capture the broad wings of the $H_\beta$ transition while excluding adjacent interference.

2. Baseline Subtraction: To isolate the emission line from the continuous background radiation, a baseline subtraction was performed. The minimum intensity value within the window was defined as the local background baseline ($I_{base} = 0$)

3. Lorentzian Fitting: Since pressure broadening in this density regime yields a profile shape dominated by the electron impact mechanism, the background-corrected experimental data were fitted with a Lorentzian function using a non-linear least-squares solver. A Lorentzian profile was chosen over a Voigt profile because the Stark width ($\Delta\lambda_{stark}$) was significantly larger than the Gaussian contributions from Doppler and instrumental broadening, rendering the Gaussian component mathematically insignificant. An example of the fitting over the raw spectrum data is shown in **Figure S1**.

4. Density Calculation: The electron density was derived from the extracted FWHM ($\Delta\lambda_{FWHM}$) using the following relation [5]:

$$n_e = 10^{23} \times \left(\frac{\Delta\lambda_{FWHM}}{4.8}\right)^{1/0.68116} \quad \text{Eq. 1}$$

Where $n_e$ is the electron density in $m^{-3}$ and $\Delta\lambda_{FWHM}$ is the full width at half maximum in nm. This calculation was performed for three experimental replicates at each spatial position, with the mean and standard deviation reported.

## 1.2. Saha-Boltzmann Method for determining Plasma temperature

While electron density provides insight into the species concentration in the plasma, the plasma excitation temperature ($T_{exc}$) serves as the primary indicator of the energetic state of the system. In this study, $T_{exc}$ was determined using the Saha-Boltzmann method, which extends the standard Boltzmann plot technique by incorporating emission lines from successive ionization stages (neutral Ar I and singly ionized Ar II).

The standard Boltzmann plot method, using only neutral lines, is often prone to large uncertainties due to the relatively small energy spread of the upper states ($\Delta E_u \sim 1 - 3eV$). By including ionic lines (Ar II), the energy range is effectively extended by the ionization potential of argon ($E_{ion} \sim 15.76\ eV$), creating a much larger lever arm for the linear regression and significantly improving the accuracy of the temperature determination.

Assuming Local Thermodynamic Equilibrium (LTE), the population densities of atomic and ionic states are coupled via the Saha-Eggert equation. The linearized Saha-Boltzmann equation is expressed as [6]:

$$\ln\left(\frac{I_{ki}\lambda}{g_k A_{ki}}\right)^* = -\frac{E_k}{k_B T_{exc}} + C \quad \text{Eq. 2}$$

where $I_{ki}, \lambda, g_k$ and $A_{ki}$ are the integrated line intensity, wavelength, statistical weight, and transition probability, respectively. The term on the left-hand side, denoted with an *, represents the "Saha-shifted" intensity. For neutral lines (Ar I), this is simply the natural logarithm of the normalized intensity. For ionic lines (Ar II), a correction term (Saha shift) is subtracted to account for the ionization equilibrium:

$$\text{Shift}_{Saha} = \ln\left(\frac{2(2\pi m_e k_B T_{exc})^{3/2}}{h^3 n_e}\right) \qquad \textit{Eq. 3}$$

This shift depends on the electron density ($n_e$), which was determined independently via Stark broadening (Section 2.4), and the temperature ($T_{exc}$) itself, necessitating an iterative solution.

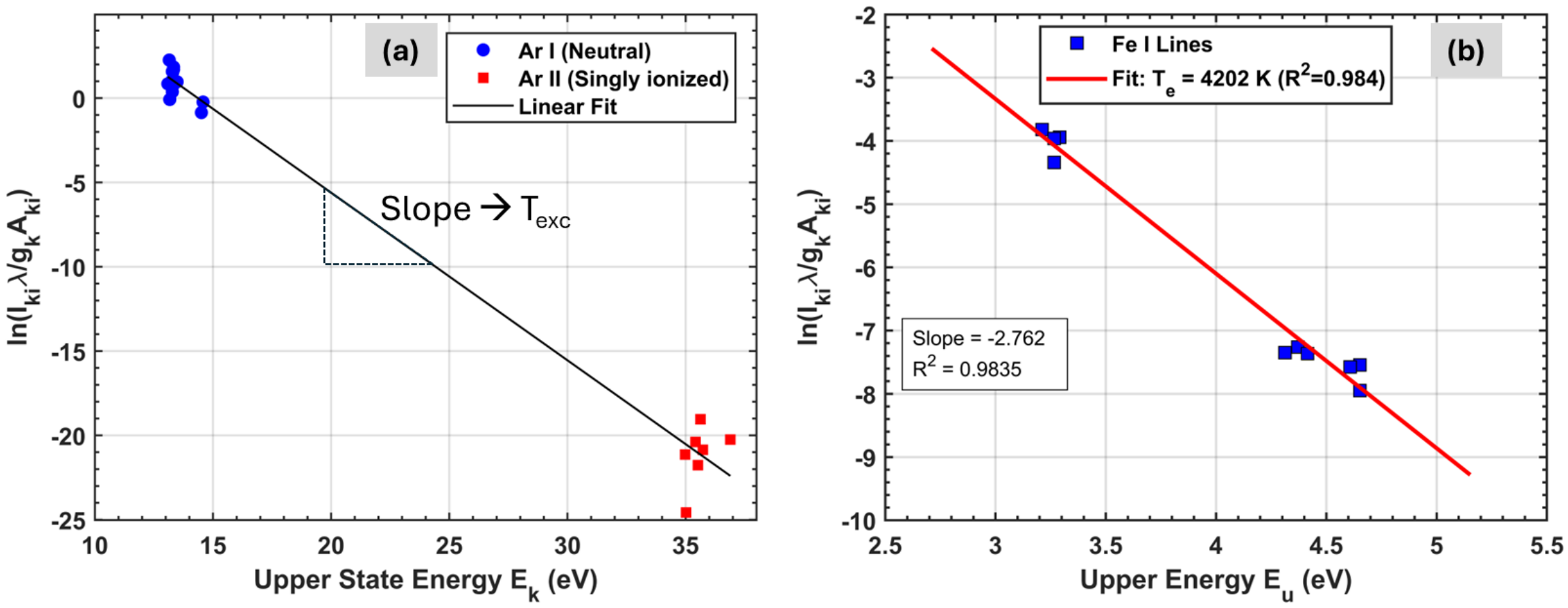

**Figure S2**: An example of the Saha-Boltzmann plot showing the linear fit of the (a) Ar I and Ar II energy and (b) Fe I lines. The slope of the linear fit gives the excitation temperature value. For Arc current I=200 A and at axial location z = 8mm

A custom iterative algorithm was developed to solve this implicit dependence and ensure the selection of high-quality spectral lines for both the Argon (Ar I/Ar II) and Iron (Fe I) datasets. The procedure, based on the method described by Aydin et al. [7], proceeded as follows:

1. Line Selection: A comprehensive set of Ar I and Ar II lines, as well as candidate Fe I lines, within the 300–800 nm range was initially selected. Atomic data ($A_{ki}, E_k, g_k$) were sourced from the NIST Atomic Spectra Database [8]. The investigated Argon and hydrogen emission lines are listed in **Table S1**.

2. Iterative Convergence: An initial temperature guess ($T_0 = 12{,}000K$) was used to calculate the initial Saha shift. A linear regression was performed on the combined Ar I/Ar II dataset to derive a new slope and temperature, as shown in **Figure S2**. This process was repeated until the temperature change between iterations fell below 1%.

3. Outlier Rejection: To mitigate errors arising from self-absorption or spectral blending, an automated optimization loop was employed. If the coefficient of determination ($R^2$) of the linear fit was below 0.98, the line with the largest residual was removed from the dataset, and the fit was re-calculated. This process continued until $R^2 \geq 0.98$ (with a minimum of 5 lines retained), ensuring that the final calculated temperature reflected a strictly thermalized population distribution fitting the Boltzmann statistic.

It is important to note that while this iterative, multi-line optimization method was utilized for calculating $T_{exc}$ from the Ar I/II and Fe I species, it was not applied to the hydrogen spectra. Within the experimental system, only three Hydrogen Balmer lines ($H_\alpha, H_\beta, and\ H_\gamma$) fall within the observable range, and the weak $H_\gamma$ emission is frequently suppressed by background continuum radiation. Because the dataset is inherently limited to these specific transitions, no iterative outlier rejection is required. Instead, the hydrogen excitation temperature was calculated directly using the classical two-line Boltzmann method based entirely on the intensity ratio of the $H_\alpha\ and\ H_\beta$ lines.

**Table S1**: Spectroscopic parameters of Argon and Hydrogen Balmer series emission lines used to calculate excitation temperature

| **Wavelength (nm)** | **$E_u$ (eV)** | **$A_{ul}$ (×$10^8$) ($s^{-1}$)** | **$g_u$** | **Species** |
|---|---|---|---|---|
| 415.859 | 14.529 | 0.014 | 5 | Ar I |
| 420.067 | 14.505 | 0.00967 | 7 | Ar I |
| 425.936 | 14.738 | 0.0415 | 1 | Ar I |
| 430.01 | 14.505 | 0.00394 | 5 | Ar I |
| 696.543 | 13.327 | 0.0639 | 3 | Ar I |
| 706.721 | 13.302 | 0.038 | 5 | Ar I |
| 714.704 | 13.282 | 0.00625 | 3 | Ar I |
| 750.386 | 13.48 | 0.445 | 1 | Ar I |
| 763.51 | 13.171 | 0.245 | 5 | Ar I |
| 811.531 | 13.075 | 0.331 | 7 | Ar I |
| 434.806 | 19.489 | 1.15 | 4 | Ar II |
| 442.6 | 19.549 | 0.742 | 4 | Ar II |
| 460.956 | 21.139 | 0.79 | 6 | Ar II |
| 476.486 | 19.868 | 0.426 | 4 | Ar II |
| 480.602 | 19.222 | 0.78 | 6 | Ar II |
| 487.986 | 19.683 | 0.79 | 4 | Ar II |
| 656.28 | 12.088 | 0.441 | 18 | Hα |
| 486.13 | 12.748 | 0.08419 | 32 | Hβ |
| 434.05 | 13.054 | 0.0253 | 50 | Hγ |

**Table S2:** Spectroscopic parameters of Fe neutral emission lines used to calculate excitation temperature

| Wavelength (nm) | $E_u$ (eV) | $A_{ul}$ ($\times 10^8$) ($s^{-1}$) | $g_u$ | Species |
|---|---|---|---|---|
| 358.119 | 4.32 | 1.04 | 11 | Fe I |
| 371.993 | 3.332 | 0.162 | 11 | Fe I |
| 373.486 | 4.178 | 0.902 | 11 | Fe I |
| 374.556 | 3.417 | 0.115 | 7 | Fe I |
| 375.823 | 4.256 | 0.634 | 7 | Fe I |
| 382.043 | 4.103 | 0.669 | 11 | Fe I |
| 385.991 | 3.211 | 0.0968 | 9 | Fe I |
| 400.524 | 4.652 | 0.264 | 7 | Fe I |
| 404.581 | 4.548 | 0.862 | 9 | Fe I |
| 406.359 | 4.608 | 0.656 | 7 | Fe I |
| 427.176 | 4.387 | 0.227 | 11 | Fe I |
| 430.79 | 4.434 | 0.338 | 9 | Fe I |
| 432.576 | 4.473 | 0.505 | 7 | Fe I |
| 438.355 | 4.312 | 0.5 | 11 | Fe I |
| 440.475 | 4.371 | 0.275 | 9 | Fe I |
| 441.512 | 4.415 | 0.117 | 7 | Fe I |
| 526.954 | 3.211 | 0.0129 | 11 | Fe I |
| 532.804 | 3.241 | 0.0115 | 7 | Fe I |
| 537.149 | 3.266 | 0.0105 | 5 | Fe I |
| 540.577 | 3.291 | 0.0093 | 3 | Fe I |
| 542.97 | 3.266 | 0.0042 | 9 | Fe I |
| 544.692 | 3.291 | 0.0036 | 5 | Fe I |

# 2. Computational Method-Supplementary information

## 2.1. Governing equations for the computational model

$$\frac{\partial \rho}{\partial t} + \vec{\nabla} \cdot (\rho \boldsymbol{V}) = 0 \quad \text{Eq. 4}$$

$$\frac{\partial}{\partial t}(\rho \boldsymbol{V}) + \vec{\nabla} \cdot (\rho \boldsymbol{V} \otimes \boldsymbol{V}) = -\vec{\nabla}\Pi + \vec{\nabla} \cdot \bar{\bar{\tau}} + \boldsymbol{J} \times \boldsymbol{B} \quad \text{Eq. 5}$$

$$\frac{\partial}{\partial t}(\rho h) + \vec{\nabla} \cdot \big((\rho h)\boldsymbol{V}\big) = \vec{\nabla} \cdot \left(\frac{\lambda}{C_p}\vec{\nabla} h\right) + \frac{5}{2}\frac{K_B}{e}\boldsymbol{J} \cdot \vec{\nabla} T + \boldsymbol{J} \cdot \boldsymbol{E} - 4\pi\epsilon_n \quad \text{Eq. 6}$$

$$\vec{\nabla} \cdot \big(-\sigma \vec{\nabla} \phi\big) = 0 \quad \text{Eq. 7}$$

$$\boldsymbol{E} = -\vec{\nabla}\phi \quad \text{Eq. 8}$$

$$\boldsymbol{J} = \sigma \boldsymbol{E} \quad \text{Eq. 9}$$

$$\vec{\nabla} \times \boldsymbol{A} = \boldsymbol{B}$$ Eq. 10

$$\nabla^2 \boldsymbol{A} = \mu_0 \boldsymbol{J}$$ Eq. 11

Here $\rho$ represents the mass density $(kg/m^3)$, $h$ is the enthalpy $(J/kg)$ $\mathbf{V} = (u, v, w)$ the velocity $(m/s)$ with $(u, v, w)$ being respective components along x, y and z directions, $\lambda$ is the thermal conductivity $(W/m/K)$ and $C_p$ is the specific heat capacity $(J/kg/K)$. Based on the low-Mach-number flow assumption, the pressure is decomposed into a thermodynamic pressure and a perturbation $(\Pi)$. The thermodynamic pressure in this work has been assumed constant at atmospheric conditions. The perturbational pressure is about 2 orders of magnitude smaller compared to thermodynamic pressure $(O(M^2)$ where Mach number M is atmost 0.1 for cases studied in this work) the term $\frac{DP}{DT}$ can thus be neglected. Furthermore, under typical discharge conditions at low magnetic Reynolds numbers, $\mathbf{E} \gg \mathbf{V} \times \mathbf{B}$, allowing the joule heating term $\mathbf{J} \cdot (\mathbf{E} + \mathbf{V} \times \mathbf{B})$ to be approximated as $\mathbf{J} \cdot \mathbf{E}$. Radiative losses $U$ are expressed as $4\pi\epsilon_N$, which depends only on local temperature and pressure. The governing equations are solved over the computational domain shown in **Figure S3**, with the corresponding boundary conditions summarized in **Table S3**. Additional details regarding the choice of boundary conditions, numerical methodology and implementation are available in our previous work [9].

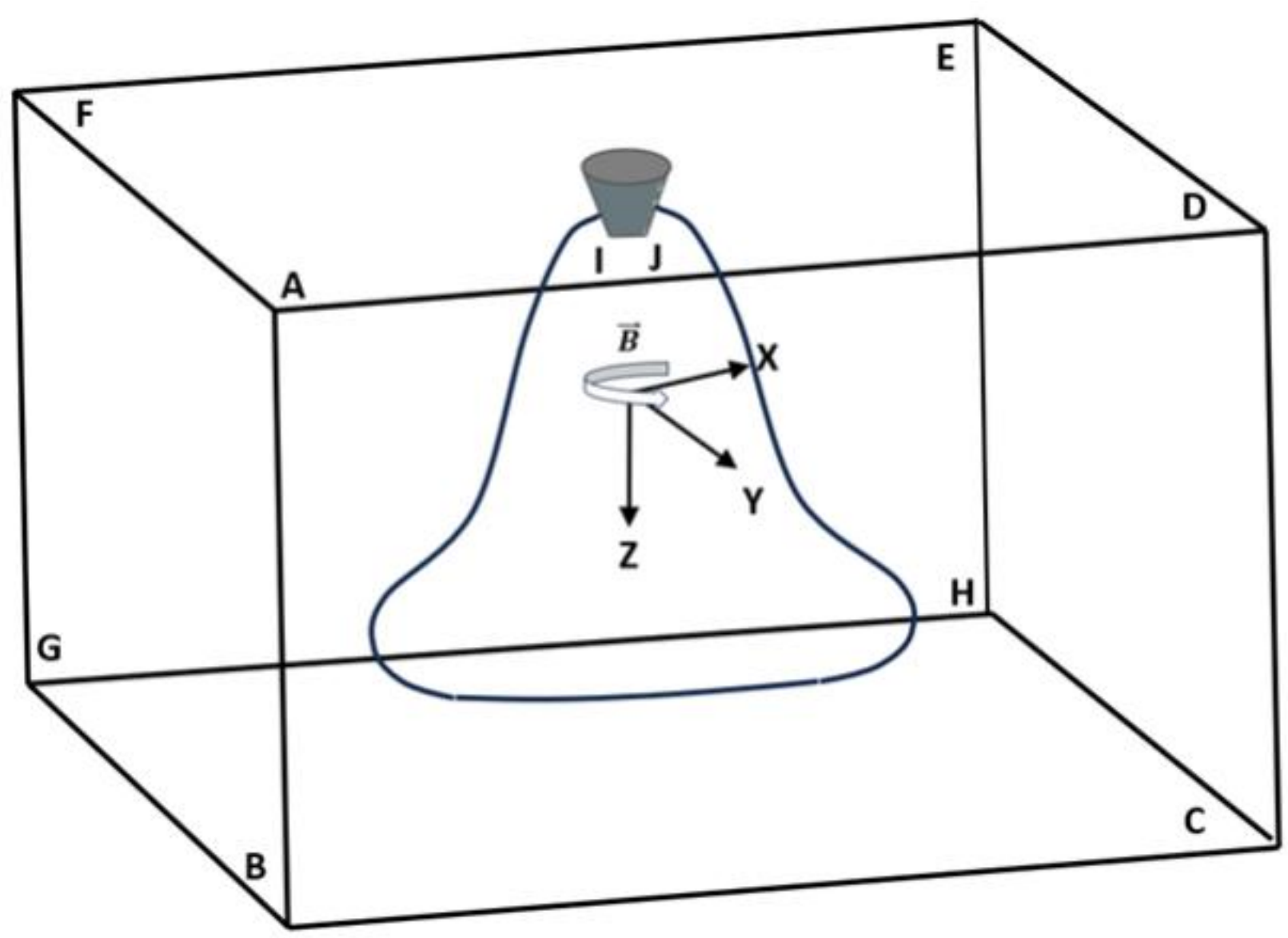

**Figure S3**: Computational domain for the arc discharge model

**Table S3**: Boundary conditions for computational model

| Field | IJ | AFGB, DEHC, ABCD, EFGH | BGHC | AFDE | Cathode body |
|---|---|---|---|---|---|
| ϕ | $\sigma \frac{\partial \phi}{\partial n} = I/\pi r_c^2$ | $\frac{\partial \phi}{\partial n} = 0$ | $\phi = 0$ | $\frac{\partial \phi}{\partial n} = 0$ | $\frac{\partial \phi}{\partial n} = 0$ |
| $T$ | $T = 3500\ K$ | $T = 1000\ K$ | $T = 1000\ K$ | $T = 1000\ K$ | $T = 1000\ K$ |
| $u$ | $u = 0$ | $\frac{\partial u}{\partial n} = 0$ | $u = 0$ | $\frac{\partial u}{\partial n} = 0$ | $u = 0$ |
| v | $v = 0$ | $\frac{\partial v}{\partial n} = 0$ | $v = 0$ | $\frac{\partial v}{\partial n} = 0$ | $v = 0$ |
| w | $w = 0$ | $\frac{\partial w}{\partial n} = 0$ | $w = 0$ | $\frac{\partial w}{\partial n} = 0$ | $w = 0$ |
| $A_x$ | $\frac{\partial A_x}{\partial n} = 0$ | $\frac{\partial A_x}{\partial n} = 0$ | $\frac{\partial A_x}{\partial n} = 0$ | $\frac{\partial A_x}{\partial n} = 0$ | $\frac{\partial A_x}{\partial n} = 0$ |
| $A_y$ | $\frac{\partial A_y}{\partial n} = 0$ | $\frac{\partial A_y}{\partial n} = 0$ | $\frac{\partial A_y}{\partial n} = 0$ | $\frac{\partial A_y}{\partial n} = 0$ | $\frac{\partial A_y}{\partial n} = 0$ |
| $A_z$ | $\frac{\partial A_z}{\partial n} = 0$ | $\frac{\partial A_z}{\partial n} = 0$ | $\frac{\partial A_z}{\partial n} = 0$ | $\frac{\partial A_z}{\partial n} = 0$ | $\frac{\partial A_z}{\partial n} = 0$ |

## 2.2. Validation of computational model and experimental results against other similar work

The fidelity of the computational approach was established against the canonical 200 A free-burning argon arc of Hsu et al.[10], whose measured isotherms the model reproduces with good agreement (**Figure S*4***). A direct experimental replication of this benchmark is not possible in the present experimental system: the electrode gap here is twice that of Hsu's 10 mm arrangement, and the anode is an iron ore, evaporating oxide melt rather than a water-cooled copper plate.

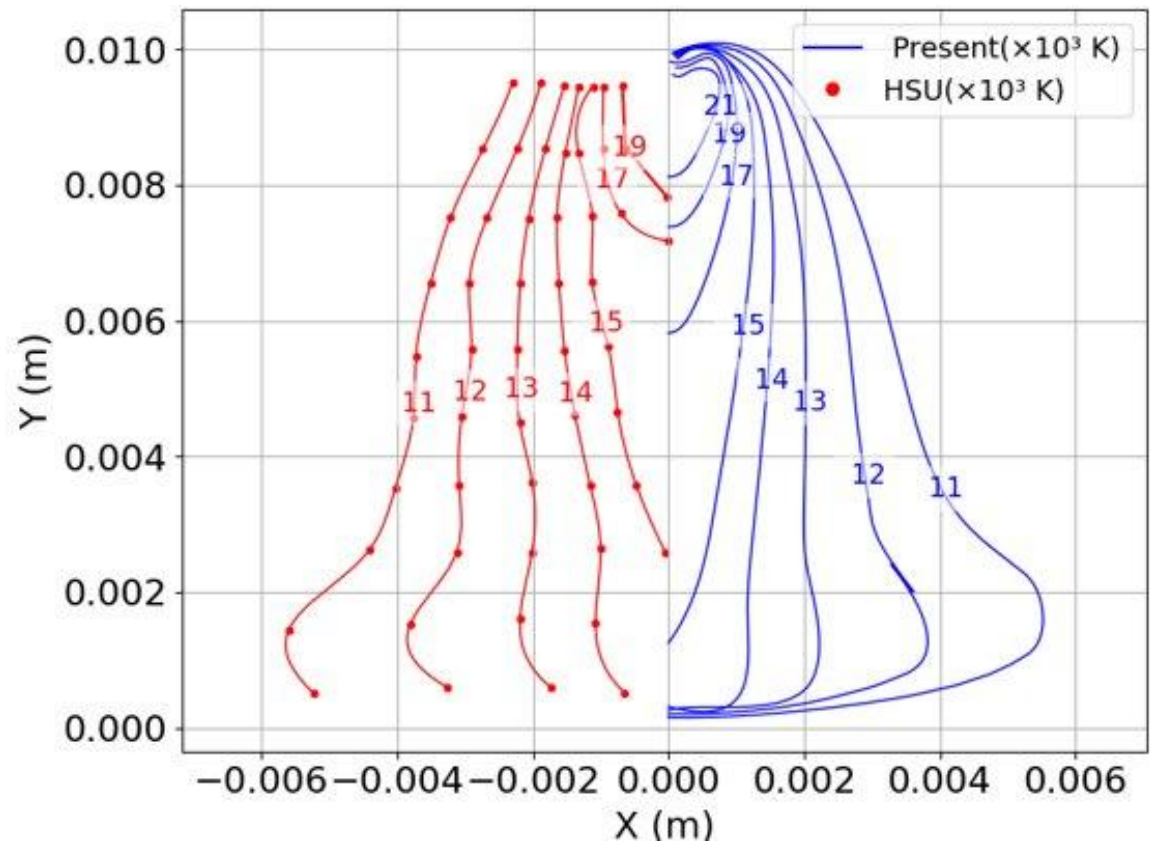

**Figure S4:** Validation of the thermal-plasma model against the canonical free-burning argon arc of Hsu et al.[10]. Computed temperature fields (blue solid lines, "Present") are shown mirrored about the arc axis (X = 0) against the reference solution of Hsu et al. (red symbols) for a 200 A pure-argon arc.